\documentclass[12pt]{article}
\usepackage{epsf} 

\begin{document}

\baselineskip 18pt

\newcommand{\sheptitle}
{Hagedorn Inflation: 
Open Strings on Branes Can Drive Inflation}

\newcommand{\shepauthor}
{
Steven Abel$^{a,}$\footnote[1]{S.A.Abel@durham.ac.uk},
Katherine Freese $^{b,}$\footnote[2]{ktfreese@umich.edu}, 
Ian I.~Kogan $^{c,}$\footnote[3]{i.kogan@physics.ox.ac.uk}}

\newcommand{\shepaddress}
{
$a$ Dept. of Mathematical Sciences, University of Durham, 
	Science Laboratories, \hfil\break
	South Rd., Durham DH1 3LE, United Kingdom\\
$b$ Michigan Center for Theoretical Physics,
        University of Michigan, Ann Arbor MI 48109-1120, USA\\
$c$ Theoretical Physics, 1 Keble Rd, Oxford OX1 3NP, UK}

\newcommand{\shepabstract}
{We demonstrate an inflationary solution to the cosmological horizon
problem during the Hagedorn regime in the
early universe.  Here the observable universe is confined to three
spatial dimensions (a three-brane) embedded in higher dimensions.
The only ingredients required are open strings on
D-branes at temperatures close to the string scale.
No potential is required.
Winding modes of the strings
provide a negative pressure that can drive inflation of our
observable universe.  Hence the mere existence of open strings on branes
in the early hot phase of the universe drives Hagedorn inflation,
which can be either power law or exponential. 
We note the amusing fact that, in the case of  stationary
extra dimensions, inflationary expansion takes place only for branes
of three or less dimensions.}

\begin{titlepage}
\begin{flushright}
MCTP-02-25
%OUTP-00-09-P\\
%hep-th/0205317\\

\end{flushright}
\vspace{0.5in}
\begin{center}
{\large{\bf \sheptitle}}
\bigskip \\ \shepauthor \\ \mbox{} \\ {\it \shepaddress} \\ 
\vspace{0.5in}
%\today
%\vspace{0.5in}
{\bf Abstract} \bigskip \end{center} \setcounter{page}{0}
\shepabstract
%\begin{flushleft}
%CERN-TH/98-375\\
\end{titlepage}

%       macros
\def\sspace{\baselineskip = .16in}
\def\dspace{\baselineskip = .30in}
\def\beq{\begin{equation}}
\def\eeq{\end{equation}}
\def\bea{\begin{eqnarray}}
\def\eea{\end{eqnarray}}
\def\bq{\begin{quote}}
\def\eq{\end{quote}}
\def\ra{\rightarrow}
\def\lra{\leftrightarrow}
\def\ups{\upsilon}
\def\bq{\begin{quote}}
\def\eq{\end{quote}}
\def\ra{\rightarrow}
\def\un{\underline}
\def\ov{\overline}
\def\ord{\cal O} 

\newcommand{\plb}[3]{{{\it Phys.~Lett.}~{\bf B#1} (#3) #2}}
\newcommand{\npb}[3]{{{\it Nucl.~Phys.}~{\bf B#1} (#3) #2}}
\newcommand{\prd}[3]{{{\it Phys.~Rev.}~{\bf D#1} (#3) #2}}
\newcommand{\ptp}[3]{{{\it Prog.~Theor.~Phys.}~{\bf #1} (#3) #2}}
\newcommand{\ijmpa}[3]{{{\it Int.~J.~Mod.~Phys.}~{\bf A#1} (#3) #2}}
\newcommand{\prl}[3]{{{\it Phys.~Rev.~Lett.}~{\bf #1} (#3) #2}}
\newcommand{\hepph}[1]{{\tt hep-ph/#1}}
\newcommand{\hepth}[1]{{\tt hep-th/#1}}
\newcommand{\grqc}[1]{{\tt gr-qc/#1}} 
\newcommand{\leqsim}{\,\raisebox{-0.6ex}{$\buildrel < \over \sim$}\,}
\newcommand{\geqsim}{\,\raisebox{-0.6ex}{$\buildrel > \over \sim$}\,}
\newcommand{\nin}{\,  \mbox{$/$ \hspace{-0.2cm} $\in$}\,}
\newcommand{\be}{\begin{equation}}
\newcommand{\ee}{\end{equation}}
\newcommand{\ba}{\begin{eqnarray}}
\newcommand{\ea}{\end{eqnarray}}
\newcommand{\nn}{\nonumber}
\newcommand{\cf}{\mbox{{\em c.f.~}}}
\newcommand{\ie}{\mbox{{\em i.e.~}}}
\newcommand{\eg}{\mbox{{\em e.g.~}}}
\newcommand{\mpl}{\mbox{$M_{pl}$}}
\newcommand{\ol}[1]{\overline{#1}}
\newcommand{\eqr}[1]{eq.(\ref{#1})}
\def\gev{\,{\rm GeV }}
\def\tev{\,{\rm TeV }}
\def\dd{\mbox{d}}
\def\etal{\mbox{\it et al }}
\def\half{\frac{1}{2}}
\def\Tr{\mbox{Tr}}
\def\bra{\langle}
\def\ket{\rangle}
\def\lim{\mbox{{\bf L}} }
\def\nlim{\mbox{{\bf NL}} }
\def\sclim{\mbox{\tiny{\bf L}} }
\def\scnlim{\mbox{\tiny{\bf NL}} }
\def\nlimc{\mbox{{\bf NL$_{closed}$}} }
\def\nlimo{\mbox{{\bf NL$_{open}$}} }
\def\Vp{V_{\parallel}}
\def\Vt{V_{\perp}} 
\def\vp{V_{\parallel}}
\def\vt{V_{\perp}} 
\def\wp{W_{\parallel}}
\def\wt{W_{\perp}} 
\def\Wp{W_{\parallel}}
\def\Wt{W_{\perp}} 
\def\Rt{R_{\perp}}
\def\ep{\varepsilon} 
\newcommand{\smallfrac}[2]{\frac{\mbox{\small #1}}{\mbox{\small #2}}}

\def\CAG{{\cal A/\cal G}}           \def\CO{{\cal O}} \def\CZ{{\cal Z}}
\def\CA{{\cal A}} \def\CC{{\cal C}} \def\CF{{\cal F}} \def\CG{{\cal G}}
\def\CL{{\cal L}} \def\CH{{\cal H}} \def\CI{{\cal I}} \def\CU{{\cal U}}
\def\CB{{\cal B}} \def\CR{{\cal R}} \def\CD{{\cal D}} \def\CT{{\cal T}}
\def\CM{{\cal M}} \def\CP{{\cal P}}
\def\CN{{\cal N}} \def\CS{{\cal S}}  

%************************************************************
%************************************************************
\section{Introduction}
\label{sec:intro}
%************************************************************
%************************************************************

	Inflationary cosmology \cite{guth} was proposed as a solution to the
horizon, flatness, and monopole problems of the standard Hot Big Bang
scenario.  The cosmological horizon problem
can be stated as follows:  The comoving size of our observable
universe today is $L_o = \int_{t_{dec}}^{t_o} {dt \over a(t)}$, where
$a(t)$ is the scale factor,
$t_{dec}$ is the time of radiation decoupling, and $t_o$ is the age
of the universe today.  This lengthscale
must fit inside the comoving size of the horizon (a causal region)
at some early time $t_n$ (before decoupling), $L_n = \int_o^{t_n}
{dt \over a(t)}$.
To explain  causal contact of all points of our observable universe
at $t_n$, we need $L_o < L_n$.  For power law expansion of the 
scale factor both before $t_n$ and after $t_{dec}$, this condition
becomes ${1 \over a_n H_n} \geq {1\over a_o H_o}$ where the Hubble
constant $H= \dot a/a$.  Inflation solves
this causality condition with a period of accelerated expansion,
$\ddot a >0$, corresponding to a superluminal expansion of the
scale factor.  In the standard 3+1 dimensional universe, the 
Friedmann Robertson Walker (FRW) equations imply
${\ddot a \over a} = - {4 \pi \over 3 m_{pl}^2} (\rho + 3p)$.
Hence accelerated expansion is provided by a negative pressure.
In standard inflationary models, this negative pressure is
provided by a vacuum energy (a potential) with $p = - \rho$.

We have found \cite{us} an entirely different source of
negative pressure:  open strings on D-branes
at temperatures close to the string scale. 
Although Einstein's equations in higher dimensions take a different
form than the FRW equation above, negative pressure can still
drive inflation. Note that there is no potential of any kind
in our model; instead, open strings on branes drive the inflation.

At sufficiently high temperatures and densities 
fundamental strings enter a curious `long string' 
Hagedorn phase~\cite{carlitz,general,deo,thorl,abkr,after}.  A
classical random walk picture can be used
to model the behaviour of the strings in 
cosmological backgrounds. The particular systems we will focus on 
are D-branes in the weak coupling limit~\cite{polch}. 
In particular, we consider the scenario in which our observable
universe is confined to three spatial dimensions (a three-brane)
embedded in higher dimensions.  We will denote the
rest of the universe outside of our 3-brane as the bulk.  We can
separate the energy momentum tensor into two components:  a 
localized component corresponding to the D-brane tension, and 
a diffuse component that spreads into the bulk corresponding to 
open string excitations of the brane. 
%in addition allow a bulk cosmological constant, although our focus in this 
%paper will be on the cosmology when the combined effect of 
%the brane and bulk cosmological constants is subdominant.
We find an interesting type of cosmological effect of a
primordial Hagedorn phase of open strings on branes: 

\begin{itemize}

\item {\it Hagedorn inflation}.  The transverse 
`bulk' components of the energy-momentum tensor can be negative. 
If all of the transverse dimensions have winding modes,
this negative `pressure' causes the brane to 
power law inflate along its length with 
a scale factor that varies as $a\sim t^{4/3}$
even in the absence of a nett cosmological constant 
(as shown in eq.(\ref{61})).
If there are transverse dimensions that are large (in the sense that 
the string modes are not space-filling in these directions),
then we can find exponential inflation (as shown in eq.(\ref{expinf})). 
{\it No potential is required here}. Merely the existence of open strings
on D-branes drives the inflation.

%\item If there is a small but negative cosmological constant,
%the Universe can enter a stable but oscillating phase; \ie
%a `bouncing' universe. The nett effect of the Hagedorn phase 
%is to soften the singular behaviour associated with the collapse. 
%Such singularity smoothing is a familiar aspect of strings,  but 
%the nice feature here is that we find it in a purely perturbative
%regime. 

\end{itemize}

We begin by discussing string thermodynamics of
open strings on D-branes at high temperatures near the string scale.
We argue that the features relevant to this paper can be derived
by obtaining the density of states from a random walk picture.
From the density of states one can derive a partition function.
Then the partition function gives the energy-momentum tensor,
the principal ingredient of Einstein's equations.  Winding
modes of the strings give rise to a negative `pressure' in the bulk
(the directions perpendicular to the three-brane on which we live).
This energy-momentum tensor, given in Eq.(\ref{eq:imp}), is
one of the most important results of the paper.
Armed with the energy-momentum tensor, we examine the resultant
cosmology. We can solve Einstein's equations with various
ans\"atze in the presence of this
negative bulk component. Our primary result is that we 
find Hagedorn inflation of our observable universe due to the negative 
pressure in the bulk.  
%In addition, we find bouncing solutions  
%with singularity smoothing behaviour.

If one assumes  adiabaticity, the inflationary growth period 
drives down the temperature of the system; eventually
the temperature drop causes
the universe to leave the Hagedorn regime, and consequently
inflation ends automatically.  In fact our solutions are only
valid for small changes in the metric corresponding to small
changes in the volumes, i.e., we can demonstrate instability
to inflationary expansion but cannot follow the solutions further.
If one uses the solutions beyond
the region of their validity, the period of superluminal
growth ends too quickly to solve cosmological problems.
Hence we do also discuss how, in non-adiabatic systems,
inflation can be sustained.

An amusing result is that power law
inflation in the universal high energy system only 
takes place if the number of large dimensions on the brane
is $p\leq 3$. In addition, branes tend to `melt' unless
$p \leq 4$. Hence one can speculate on the role of these
effects in the fact that our observable universe has three
large dimensions.

\section{The Hagedorn phase and random walks} 
 
Previous work ref.\cite{abkr} obtained the thermodynamic properties
of D-branes in toroidal compactification using both
microcanonical and canonical ensembles.  The same results
can be obtained from  a random walk argument, where the
thermodynamic properties are independent of the details of the
compactification (e.g. nontoroidal) and the degree of supersymmetry.
Here we
will derive an expression for the partition function
based on a random walk argument.
The reader interested only in the result for the energy-momentum
tensor and the resultant cosmology should proceed directly
to Eq.(\ref{eq:imp}).

\subsection{String thermodynamics and the Hagedorn phase}

The Hagedorn phase arises in theories containing fundamental
strings because they have a large number of internal degrees of 
freedom. Indeed, because of the existence of many 
oscillator modes, the density of states 
grows exponentially with energy $\varepsilon$, $\omega(\varepsilon) 
\sim \varepsilon^{-b}
e^{\beta_H \varepsilon}$, where the inverse Hagedorn temperature $\beta_H$ 
(where $\beta =1/T$)
and the exponent $b$ depend on the particular
theory in question (for example heterotic or type II)~\cite{carlitz}. 
For type I,IIA,IIB strings the numerical value of the 
inverse Hagedorn temperature
is $\beta_H=2 \sqrt{2} \pi$ in string units.
It is easy to see that thermodynamic quantities, such as the 
entropy, are liable to diverge at the 
Hagedorn temperature;
obtaining the partition function $Z$ with the
canonical ensemble and multiplying
by the usual Boltzmann factor $e^{-\beta \varepsilon}$,
one finds an integral for the partition function (for large $\varepsilon$)
$Z \propto \int d\varepsilon \varepsilon^{-b} e^{(\beta_H - \beta) 
\varepsilon}$ which diverges at
$\beta = \beta_H$ for $b \leq 2$. 

In the Hagedorn regime, fundamental strings can be described
as `long strings' in a random walk analogy.  [In a one-dimensional
random walk, with each step one has 2 choices of direction
(right or left) leading after `n' steps to the factor $2^n = e^{(log 2)n}$,
which mimics the exponential behavior of $\omega(\varepsilon)$ above.]
The size of the random walk (the distance covered by the string)
is given by the length scale $\sqrt{\varepsilon}$. The physical setup
is shown in Figure 1.  

\begin{figure}[htp]
\begin{center}
%\vspace*{1in}
\hspace*{0.5in}
\epsfxsize=5.0in
\epsffile{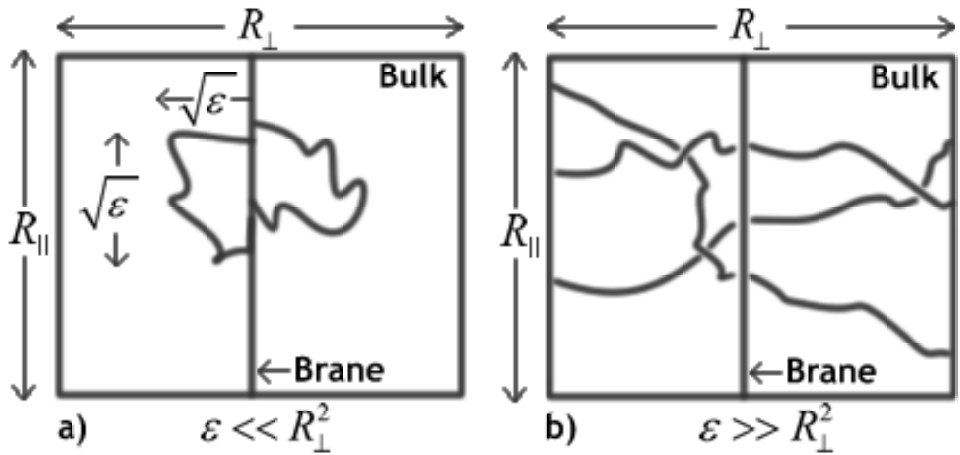}
%\vspace{1in}
\caption{We consider a volume, portrayed by the box,
containing a single brane embedded in the bulk.
The quantity $R_{||}$
indicates the size of the dimensions parallel to the brane 
(for cosmology, $R_{||} \rightarrow \infty$) and $R_{\perp}$
indicates the size of the dimensions perpendicular to the brane.
The random walk of strings attached to the brane 
traverses a distance $\sqrt{\varepsilon}$.
Figs. 1a represents the case $\sqrt{\epsilon} << R_{\perp}$.
Fig. 1b represents the case  
$\sqrt{\epsilon} >> R_{\perp}$.  This latter case is our standard
high energy case in which all modes are space-filling
(effectively, ``winding modes'').   }
\end{center}
\end{figure}

\vfill\eject
We consider a volume, portrayed by the box,
containing a single brane.  The quantity $R_{||}$
indicates the size of the dimensions parallel to the brane 
(for cosmology, $R_{||} \rightarrow \infty$) and $R_{\perp}$
indicates the size of the dimensions perpendicular to the brane.
The random walk of strings attached to the brane 
traverses a distance $\sqrt{\varepsilon}$.
Figs. 1a and 1b show the cases $\sqrt{\epsilon} << R_{\perp}$ and 
$\sqrt{\epsilon} >> R_{\perp}$ respectively.  
In this paper we are particularly
interested in the high energy case of fig. 1b, in which the strings
fill all the space; for the case of toroidal compactification
this corresponds to winding modes.  We will use the nomenclature
`winding modes' to encompass these space-filling modes regardless
of the type of compactification.  We define $d_o$ to be the number
of dimensions transverse to the brane in which there are {\it no windings}.
As our standard `high-energy' regime, we will take the case of $d_o = 0$, 
so that all dimensions have windings;
this is the system which is always reached provided that the 
the energy density is high enough.
%These systems are equivalent to $D-1$ branes
%where $D$ is the total number of space-time dimensions
%(in other words freely moving open strings). (Once there are many 
%windings, we can 
%T dualize the Dirichlet directions so that they become Neumann directions
%much smaller than the string scale -- the winding modes become 
%a spectrum of Kaluza-Klein modes indicating
%open strings which are energetic enough to probe all of the $D-1$ Neumann
%dimensions -- even the small ones.)

In ref.\cite{abkr}, results for the entropy density
for various limiting systems of different values of $d_o$
were obtained in the {\em microcanonical ensemble}
working in an approximation to the thermodynamic limit. 
%n this approach one can calculate the temperature
%of the subsystems from 
%\ba
%S_i(E_i)&=&\log \Omega_i(E_i)\nn\\
%T_i^{-1}&=& \beta_i = \partial S_i /\partial E_i  .
%\ea
%where $\Omega_i$ is the microcanonical density of states
%in the subsystem $i$. 
In ref.\cite{abkr} a 
random walk interpretation with the appropriate combinatorics
was also used to obtain
 the distribution function $\omega (\ep)$ for
open strings attached to 
a brane:
\be
\omega (\ep)_{\rm open} \sim {\Vp \over  
\ep^{d_\perp /2}} \;\exp\,(\beta_H\,\ep) \,\,\hspace{1cm} 
R_{\perp} \gg \sqrt{\ep} \, .
\ee 
and
\be
\label{eq:flatspace}
\omega(\ep)_{\rm open} \sim {\Vp \over \Vt}  
\;\exp\,(\beta_H \,\ep) \,\,\hspace{1cm} 
R_{\perp} \ll \sqrt{\ep} .
\ee 
%n agreement with eq.(\ref{omega}). 

%\subsection{Random walks in a cosmological background}

One can now adapt the random walk to a cosmological background.
In the case of a non-trivial 
metric the most natural interpretation of the parameter $\ep$ is that it is 
the proper length of 
a string in the bulk and certainly we can always go 
to the local inertial frame in which a small portion of the string 
has the usual Euclidean energy~$\equiv$~length equivalence. 

We first make the usual quasi-equilibrium 
approximation that equilibrium is established much more 
quickly than 
any change in the metric so that 
the metric may be taken to be approximately constant
in time when evaluating properties such as density.
In order to simplify matters, we also assume
that the metric is expressed in terms of parallel dimensions $x$ and 
transverse ones $y$  
\be 
ds^2 = -n^2 dt^2 + g_{||ij}dx^idx^j +g_{\perp nm}dy^ndy^m ,
\ee
with the brane lying at $y^n=0$.
We also define
an averaging over the extra dimension with an overbar, 
\be 
\label{eq:avgquant}
\overline{O} = \frac{\int dy \sqrt{g_\perp} O(y) }{\int dy'
\sqrt{g_\perp } }\, .
\ee
In \cite{us}, we found
the density of states for the limiting systems
% with critical
%exponent $\gamma$, {\bf L}[$\gamma$].  
%For example, in our `standard' high density, high energy regime with $d_o=0$
%(windings in all transverse dimensions),
%we found the density of states to be
%\be 
%\omega(\varepsilon ) 
%\sim 
%\frac{ \overline{\Vp }}{\overline{\Vt}} e^{\beta_H\varepsilon},
%\ee
to be the same expression as in the flat space case of eq.
(\ref{eq:flatspace})
but with all volumes averaged over transverse dimensions as in 
eq.(\ref{eq:avgquant}).
% and (\ref{eq:totalperp}). 

\subsection{Partition Function}

Now that we have obtained a density of states of open strings attached
to branes near the Hagedorn temperature, we can find
the partition function at temperature $1/\beta$,
\be
\log Z(\beta, \overline{\vp},\overline{\vt}) \sim \int d\varepsilon
\,\, \omega(\varepsilon) \,\, e^{-\beta \varepsilon} .
\ee
In the standard $d_o=0$ case,
\be
\label{eq:partition}
\log Z(\beta,\overline{\Vp},\overline{\Vt} ) = 
2 \frac{\overline{\Vp}^2 \beta_H^2}{\overline{\Vt} (\beta^2-\beta_H^2) } 
+ {\rm nonsingular} \,\, {\rm cutoff}\,\, {\rm terms} 
%a_c \overline{\Vp} - \overline{\vp} 
%\rho_c \frac{(\beta^2-\beta_H^2)}{2\beta_H} ,
\ee
%where $a_c$ and $\rho_c$ are a critical pressure and 
%energy density (defined with reference to the brane dimensions, 
%\ie with dimensions $E_c/\overline{\Vp}$)
%which are of order unity in string units~\cite{abkr}.
%Here, subscript-$c$ refers to critical quantities to remain in the
%Hagedorn phase.
which are the successive terms in a saddle point approximation.
In \cite{us}, we present the density of states and partition
function for arbitrary values of $d_o$ as well.

\section{Stress-energy tensor $T_{\mu\nu}$ in a bulk Hagedorn phase}

We now use these thermodynamic results to 
find the bulk energy momentum tensor during the Hagedorn regime.
We may find the energy momentum tensor from 
\be 
\label{eq:tuv}
\langle T^\mu_\nu \rangle = 2 \frac{g^{\mu\rho}}{\sqrt{g}} 
\frac{\delta \log Z(\beta,\overline{\Vp},\overline{\Vt}) } 
{\delta g^{\rho\nu}}.
\ee  
%where $\beta,~\overline{\vp},~\overline{\vt}$ are given by 
%eq.(\ref{eq:betah}).,
%(\ref{eq:parperp}),(\ref{eq:avgquant}),(\ref{eq:totalperp}).
We will treat the functional derivative with respect to $g_{\mu\nu}$
in the following way.  We assume that small changes in the metric 
correspond to making small changes in the volumes in $\log Z$, 
\eg for a single extra dimension
\be
\label{eq:fnal}
{\delta Z \over \delta g_{55}} = \int dx' {\delta Z \over \delta
\overline{V}_\perp(x')} 
{\delta \overline{V}_\perp (x') \over \delta g_{55}} \, .
\ee
Then, in the case of only one extra dimension, we can write 
\be
\label{eq:fnal2}
\overline{\Vt} = \int dy \sqrt{g_{55}} =
{\int d^5x \sqrt{g_{55}} \over \int d^4x}
= {1 \over v\beta} \int d^5x \sqrt{g_{55}} \, ,
\ee
and 
\be 
\label{eq:fnal3}
{\delta \overline{\Vt} \over \delta g_{55}} = \frac{1}{  2
\sqrt{g_{55}}  \Vp \beta} \, .
\ee
Then from eqs.(\ref{eq:fnal}--\ref{eq:fnal3}) we can determine
the functional derivative in eq.(\ref{eq:tuv}).
Our ansatz automatically means that $T_{05} = 0$
and hence $G_{05} = 0$; in other words we are not considering energy exchange
between the brane and the bulk. (In general there might
be energy flux between the two.)

We now summarize the results for the energy-momentum tensor.
We first drop the overline notation of the previous section 
and simply redefine $\Vt$ and $\Vp$ to be the transverse and parallel 
volumes covariantly averaged over the region
of the transverse dimensions covered by the strings.
%(See eqs.(\ref{eq:parperp}),(\ref{eq:totalperp}),(\ref{eq:ep}).)
We define an energy density $\rho$ of strings 
\be 
\rho = \frac{E}{\Vp}
\ee
and define
\be
\label{eq:defgamma}
\gamma = d_o/2-1
\ee  
where $d_o$ is the number of dimensions with no windings.
%For the limiting systems with critical exponent $\gamma$, {\bf L}[$\gamma$], 
We find that the `bulk' components of 
the energy momentum tensor are given by
\ba
\label{eq:imp}
\hat{T}^0_0 & = & -\hat{\rho}
\nonumber \\
\hat{T}^i_i & = & \hat{p}_{\gamma} \nonumber \\
\hat{T}^m_m \equiv p_{bulk} &\approx & \left\{
\begin{array}{ll}
-\hat{p}_{\gamma} & \mbox{transverse with windings}
\\
0 & \mbox{transverse without windings,}
\end{array}\right.
\label{emgamma}
\ea
where $\hat{\rho}=\rho/\Vt$ and 
\be
\label{eq:phat}
\hat{p}_{\gamma} =   
\left\{
\begin{array}{ll}
\frac{1}{\Vt^{3/2}} 
\rho^{\frac{\gamma}{\gamma-1}}
\;\;\;\;\;\;&{\rm if}\;\;\;\;{\gamma=-1,-\half,\half}
  \nn\\
\;\nn\\
\frac{1}{\Vt^{3/2}} 
\log\rho
\;\;\;\;\;\;&{\rm if}\;\;\;\;{\gamma=0}
\nn\\
\,\nn \\
\frac{1}{\Vt^{3/2}} 
e^{-\rho}
\;\;\;\;\;\;&{\rm if}\;\;\;\;{\gamma=1,}
\end{array}
\right.
\ee
wherever there are strings present, and zero otherwise.
In particular, for our standard high energy case of $\gamma = -1$,
the negative bulk pressure is:
\be
\label{eq:pbulk}
p_{bulk} \sim - \rho^{1/2}/\Vt^{3/2} .
\ee
%The approximation in the transverse components 
%is valid when the parallel volume changes only by a small fraction 
%over the transverse directions. 

As expected $\hat{T}_0^0$ resembles the local energy density
of strings. The  $\hat{T}^i_i$ represents a relatively small 
pressure coming from Kaluza-Klein modes in the Neumann directions and 
$\hat{T}^m_m$ is a {\em negative} pressure coming from winding modes
in the Dirichlet directions. If we T-dualize the Dirichlet
directions these `winding modes'  
also become Kaluza-Klein modes in Neumann-directions and $T_m^m$ becomes 
positive. Thus negative $T_m^m$ reflects the fact that we have 
T-dualized a dimension much smaller than the string scale thereby 
reversing the pressure. For this reason negative $T_\mu^\mu$ is 
expected to be a 
general feature of space-filling excitations in transverse dimensions. 
The most important result of this section is the negative bulk
pressure found in Eqs.(\ref{emgamma}) and (\ref{eq:pbulk}).

%Those who are familiar with T-duality may suspect an apparent conflict 
%with this reversal of pressure when we T-dualize the extra dimensions. 
%However we stress that T-duality is 
%maintained. Pressure is merely 
%{\em defined} as the change in free energy with 
%{\em increase} in volume. Thus since free energy is invariant under
%T-duality the pressure must reverse sign. On the other hand the 
%cosmological consequences in the T-dual system must of course be the same. 

\section{Cosmological Equations in $D=5$}

%\vspace{0.5cm}
%\noindent\underline{ {\it Metric and Einstein Equations}}
%\vspace{0.5cm}

We now consider
a D-brane configuration that has 3 large parallel 
dimensions (\ie the `observable
universe') and only one transverse dimension $y$ that supports winding 
modes.
%; we expect to find the 
%same behaviour in higher dimensions. 
For now, we also assume adiabaticity.

We use the metric, 
\be 
\label{simple}
ds^2 = -n^2 dt^2 + a^2 d{\bf x}^2 + b^2 dy^2 ,
\ee
which foliates the space into
flat, homogeneous, and isotropic spatial 3-planes.
Here ${\bf x} = x_1,x_2,x_3$ are the coordinates on the
spatial 3-planes while $y$ is the  coordinate
of the extra dimension.  For simplicity, we make a further
restriction by imposing $Z_2$ symmetry under $y
\rightarrow -y$.  
Without any loss of generality we choose the 3-brane
of the `observable universe' to be fixed at $y=0$.
%(Even if the brane is moving we can always reparameterize the
%theory so that the 3-brane is fixed in coordinate space, although
%in doing so we lose further freedom of gauge choice~\cite{chuf}.)
%With our assumption of homegeneity 
We can associate a scale 
factor with the parallel dimensions $a(t,y)$ and one for the 
`extra' transverse dimensions $b(t,y)$.
We define $a_0(t(\tau))=a(t,0)$ 
as the scale factor describing the expansion
of the 3-brane where $t(\tau) \equiv \int d\tau n(\tau,y=0)$ is the
proper time of a comoving observer.

Several authors \cite{KKOP} - \cite{yetmorerefs}  have
presented the bulk Einstein equations.
% but for completeness we 
%briefly restate the results; 
For illustrative purposes we restate the results for the 55 (yy) equation:
%\ba
%\hat{G}_{00} &=& 3\Biggl\{\frac{\dot{a}}{a}\,\Biggl(\frac{\dot{a}}{a} +
%\frac{\dot{b}}{b}\Biggr) -\frac{n^2}{b^2}\,\Biggl[\frac{a''}{a} +
%\frac{a'}{a}\,\Biggl(\frac{a'}{a} - \frac{b'}{b}\Biggr)\Biggr]\Biggr\}
%= \hat{\kappa}^2 \, \hat{T}_{00}\,,\label{00}\\[4mm] 
%%
%\hat{G}_{ii} &=& \frac{a^2}{b^2}\Biggl\{\frac{a'}{a}\,
%\Biggl(\frac{a'}{a} + 2\frac{n'}{n}\Biggr) -\frac{b'}{b}\,\Biggl(\frac{n'}{n}
%+2\frac{a'}{a}\Biggr) +2 \frac{a''}{a} +\frac{n''}{n}\Biggr\}\nonumber\\[4mm]
%&+& \frac{a^2}{n^2}\Biggl\{\frac{\dot{a}}{a}\,\Biggl(-\frac{\dot{a}}{a} +
%2\frac{\dot{n}}{n}\Biggr) -2\frac{\ddot{a}}{a}+ \frac{\dot{b}}{b}\,
%\Biggl(-2\frac{\dot{a}}{a} + \frac{\dot{n}}{n}\Biggr) -
%\frac{\ddot{b}}{b}\Biggr\}= \hat{\kappa}^2\,\hat{T}_{ii}\,,
%\label{ii}\\[4mm]
%%
%\hat{G}_{05} &=& 3\Biggl(\frac{n'}{n} \frac{\dot{a}}{a}
%+ \frac{a'}{a} \frac{\dot{b}}{b} -\frac{\dot{a}'}{a}\Biggr)
% = \hat{\kappa}^2\,\hat{T}_{05}\,,
%\label{05}\\[4mm]
%%
\be
\hat{G}_{55} = 3\Biggl\{\frac{a'}{a}\,\Biggl(\frac{a'}{a} +
\frac{n'}{n}\Biggr) -\frac{b^2}{n^2}\,\Biggl[\frac{\dot{a}}{a}\,
\Biggl(\frac{\dot{a}}{a}-\frac{\dot{n}}{n}\Biggr) +
\frac{\ddot{a}}{a}\Biggr]\Biggr\} = \hat{\kappa}^2\,\hat{T}_{55}\,, 
\label{55}
\ee
%\ea
%%%%%%%%%%%%%
where $\hat{\kappa}^2=8\pi \hat{G}=8\pi/M_5^3$,
$M_5$ is the five-dimensional Planck mass, and the dots and primes
denote differentiation with respect to $t$ and $y$, respectively.
As stated earlier, our ansatz implies that 
$T_{05}=0$ in the bulk.  Shortly we will use the $T_{\mu\nu}$
appropriate to the Hagedorn regime on the right hand side of
Einstein's equations.  

From Eqn.(\ref{55}),
one can see right away
the importance of negative components of the bulk energy momentum tensor.
In the case where time derivatives of $a$ dominate on the left
hand side of the equation, 
one has ${b^2 \over n^2} \Biggl[ \Biggl({\dot a \over a} \Biggr)^2
+ {\ddot a \over a} \Biggr] = - \hat{\kappa}^2\,\hat{T}_{55}$.
One can see that accelerated expansion $\ddot a >0$, which
is required for inflation, takes place with a negative $\hat{T}_{55}$
such as that found in Eqns.(\ref{emgamma}) and (\ref{eq:pbulk}). 

%************************************************************
%************************************************************

%\vspace{0.5cm}
%\noindent\underline{ {\it Boundary conditions}}
%\vspace{0.5cm}

%************************************************************
%************************************************************

In addition to the bulk Einstein equations, we have boundary
conditions (the Israel jump conditions ~\cite{KKOP}-\cite{yetmorerefs})
due to the fact that our observable brane is embedded in the bulk.
We will assume that
the energy momentum tensor on the boundary can be written in a
perfect fluid form, 
%\begin{equation}
%\label{eq:energydensity}
$t^0_0 = \rho_{br}$
%\end{equation}
and 
%\begin{equation}
%\label{eq:pressure}
$t^1_1= - p_{br}$,
%\end{equation}
where $\rho_{br}$ and $p_{br}$ are the energy density and
pressure, respectively, measured by a comoving observer.
For the metric in eq.(\ref{simple}) and 
the brane at a $Z_2$ symmetry fixed plane, 
the Israel conditions become
\ba
3[a'/a]_0 &=& -\hat{\kappa}^2 b_0 \rho_{br} \nn\\
3[n'/n]_0 &=& \hat{\kappa}^2 b_0 (2\rho_{br}+3 p_{br}) \, .
\ea
In the next section, we will solve Einstein's equations together
with the constraints provided by the Israel conditions,
using the energy momentum tensor we have obtained for the 
primordial Hagedorn regime.  

Before turning to our results, let us compare the relative
%It is interesting at this point to consider the relative 
contributions of the stringy excitations and of the D-brane tension
to the cosmology. Somewhat counterintuitively, we conclude
that the diffuse stringy component can have a dominant effect
on the cosmology even in the weak coupling limit (see the discussion
in \cite{us}).  The intrinsic tension of the D-brane, 
$T_D \sim 1/\hat{\kappa}$, with $\hat\kappa \sim g_s$,
appears large for small coupling.
%
%To see this, first note that the effective gravitational 
%coupling $\hat{\kappa}$ is given in terms of the string coupling $g_s$ by 
%\be 
%\frac{1}{\hat{\kappa}^2} \sim
%\frac{m_s^{D-2} Vol_{D-5} }{g_s^2} \sim M_5^3 .
%\ee
%where $Vol_{D-5}$ is the volume of the compactification from $D=10$ to
%$5$ dimensions, $m_s$ is the string scale and $M_5$ is the 
%5-dimensional Planck mass. 
%The intrinsic tension of a D-brane is given by 
%\be 
%\label{bbb}
%T_{Dp}=\frac{1}{(2\pi)^p g_s} \sim \frac{1}{\hat{\kappa}\sqrt{Vol_{D-5}}}
%\ee
%For convenience we will henceforth assume that $ Vol_{D-5}\sim m_s^{-5}$ 
%so that $g_s$ and $\hat{\kappa}$ are of the same order of
%magnitude in string units (where $m_s\sim 1$). 
%(This need not be the case if some of the other space dimensions
%are compactified with a radius much larger or smaller than  
%the string scale.)
%Thus the intrinsic tension of the D-brane
%satisfies $\rho_{br} \sim 1/\hat{\kappa}$,
%and in the small coupling limit one might expect it to 
%dominate the cosmology.  However, this is not the case.
%If we substitute the Israel matching conditions 
%into the cosmological equations (\ref{00}-\ref{55}), 
%we see that the contributions of the 
%D-brane tension and of the stringy components in the cosmological 
%equations are of order  
However, using the Israel conditions, we see that the contribution
of the brane tension to terms in Einstein's equations is of the 
form $\hat \kappa^4 T_D^2 \sim \hat\kappa^2$ while the contribution
of the stringy components is $\hat\kappa^2 \rho/\Vt$ or $\hat\kappa^2
p_\gamma$.  Since we have assumed 
$\rho \leq T_D$ in order for the perturbative
treatment of the D=brane to be valid,
%\[
%{\hat{\kappa}}^4 \rho_{br}^2 \mbox{\hspace{0.3 cm }and \hspace{0.3 cm }}
%{\hat{\kappa}}^2 \frac{\rho}{\vt} \mbox{ ~,~} {\hat{\kappa}}^2\hat{p}_\gamma
%\]
%respectively. When the density of the string gas 
%is close to the critical Hagedorn density, $\rho_c\sim 1$ (the effective 
%lower bound), all the terms are of order $\hat \kappa^2 $ and are therefore 
%comparable. Since the density of the string gas can be
%significantly larger than the lower bound of $\rho_c \sim 1$, 
%clearly the bulk stringy contribution can be significantly
%larger than the brane contribution.
%
%We can further see the dominance of the bulk stringy component
%in the weak coupling limit.  All our work has assumed that
%\be 
%\label{bound}
%\rho \leqsim T_{Dp} .
%\ee
%If this condition is violated, 
%the thermal energy of the brane is larger than 
%its rest mass and one would expect our perturbative treatment of the 
%D-brane to break down (see ref.\cite{abkr} for possible outcomes.)
%(Note that this bound does not apply {\it{directly} }
%to orientifolds which are stuck at fixed points.)
%Since $T_{Dp} \propto {1 \over \hat{\kappa}}$,
%we can see that the amount of thermal energy that can be
%loaded onto the brane increases as the coupling gets weaker.
%When this bound is saturated, 
we see that the stringy contribution to the cosmological 
equations can be ${\cal O}({\hat{\kappa}})$ while 
that from the intrinsic tension is ${\cal O}
({\hat{\kappa}}^2)$; thus weak coupling 
is advantageous for string dominance in the cosmology.
In addition, the Yang-Mills degrees of freedom on the brane
are subdominant to the bulk degrees of freedom near $T_H$.

\section{Results: Behaviour of scale factors $a(t)$ and $b(t)$:
Inflation without Inflatons}

Our results are obtained by solving 
Einstein's equations
together with the constraints provided by the Israel conditions. 
We use the energy momentum tensor derived above in Eq.(\ref{eq:imp})
 appropriate to a primordial Hagedorn epoch.
First we will consider the case where
the brane is energy/pressureless and that $\dot{b}=0$
(where $b$ is the scale factor of the extra dimension).
We then catalogue more general types of behaviour that
can occur for $\dot{b}\neq 0$.

\subsection{Case I: 
$\Lambda_{br} = \Lambda_{bulk} =\dot{b}=0$}

Let us first impose
$\dot b=0$ simply as an external condition; \ie
the extra dimension is fixed in time.  
We also set the cosmological constants  both on the 
brane and in the bulk to zero, 
$\Lambda_{br} = \Lambda_{bulk}= 0$. All other 
contributions to the energy momentum localized on the brane 
(for example massless Yang-Mills degrees of freedom)
have an entropy that is subdominant 
to the limiting bulk degrees of freedom as long as the string
energy density 
$\rho> \rho_c$ (where $\rho_c$ is a critical Hagedorn density of order 1
in string units). 
For the moment we will therefore neglect the brane energy-momentum
and set $a'(t,0)=n'(t,0)=0$.
In this discussion the brane at $y=0$ 
is playing no role in determining the evolution of the
cosmology; the scale factor $a_0$ changes purely as a result of the 
bulk equations. 
%\vspace{0.5cm}
%\noindent\underline{ {\it A. Solutions to the $T_{55}$ equation}}
%\vspace{0.5cm}
It is simple to solve the 55 equation
for the scale factor $a_0(t) \equiv a(t,0)$ on our brane (at $y=0$).
%\be
%$a(t,0)=a_0(t)$
%\ee
%for all the {\bf L}$[\gamma]$ systems.

\noindent\underline{ {\it Standard case with $d_o=0$:} }
In the high energy case with windings
in all transverse dimensions we 
see from Eqs.(\ref{eq:imp}) and (\ref{eq:pbulk})
that $T_5^5 \sim \sqrt{\rho} \sim a^{-3/2}$
(where, again, $\rho$ is the local energy density measured with 
respect to the volume $\Vp$)
and hence, 
\begin{equation}
\label{61}
a_0(t) \sim t^{4/3} \, ,
\end{equation}  
\ie {\em power law inflation}.
For the more general case of `p' large parallel dimensions
in any number of extra perpendicular dimensions, we find
$T_5^5 \sim a^{-p/2}$ and $a \sim t^{4/p}$.  
Here adiabaticity has been assumed in taking $E \sim S \sim {\rm const}$.
We note the amusing
result that inflation requires $p \leq 3$ (for our brane
$p=3$) and speculate on its role in our
three large extra dimensions.

\noindent\underline{ {\it Case with $d_o=2$:} }
In the case of 2 dimensions with no windings,
we find a period of  
{\em exponential inflation}:
%. To find this solution we 
%write $a_0^2=\exp F$ and the $T_5^5$ equation becomes 
%\[
%\ddot{F} +\dot{F}^2 = \frac{2 \hat{\kappa}^2}{3} 
%\left( \log{\rho(0)} - \frac{3}{2}F \right)
%\]
%which may easily be solved to give 
\be 
\label{expinf}
\frac{a_0(t)}{a_0(0)} = \exp \left( 
{
-\frac{t^2\hat{\kappa}^2}
{8\Vt^{\frac{3}{2}}} +t \sqrt{\frac{\hat{\kappa}^2}{6 
\Vt^{\frac{3}{2} }} \log (\rho(0)e^{3/4})}  }\right) 
\ee
where $\rho(0)$ is the initial density at $t=0$. 
Initially the second term in the exponent dominates
and there is exponential inflation.
This solution has an automatic end to inflation 
%when 
%$da_0(t)/dt \sim 0$, \ie when the two terms in the
%exponent are roughly of the same magnitude. Inflation ends at
%time $t \sim\hat{\kappa}^{-1}\sqrt{\Vt^{3\over 2} \log\rho (0)}$
%which corresponds to $\log \frac{a_0^3(t) }{a_0^3(0)} \approx \log \rho(0)$.
%During a period of adiabaticity, the entropy and energy density
%dilute as $1/a_0^3$ because the Hagedorn phase is always 
%to a first approximation like pressureless matter 
%($\hat{p}_{-1}\ll \hat{\rho}$). Hence we find that the above condition
%for inflation to end happens 
when $\rho(t) \sim 1$ (in string units), just 
as the system is dropping out of the Hagedorn phase. 

\noindent\underline{ {\it  Other systems:}} 
We summarize the energy momentum tensors and cosmological 
behavior of $a_0(t)$ for systems with $d_o$
perpendicular dimensions without windings
in Table 1, where $p=3$ in our `toy-model'. 
Superluminal expansion is found for systems with $d_o \leq 2$,
i.e., no more than two perpendicular dimensions without windings.
%In all these cases it should be noted that 
%during a period of adiabaticity, the entropy density 
%dilutes as $1/a_0^3$. The reason for the 
%various different types of behaviour is of course due to 
%$T_5^5$ which may drop off more slowly than $1/a_0^3$
%and in the {\bf L}[0] case drops only logarithmically with the expansion. 
%Note inflationary (superluminal) expansion for $\bf{L}$[$\gamma \leq 0]$.

In adiabatic systems the Hagedorn regime and hence the inflationary 
behaviour we have found eventually come to an end.  
For a system to be in the Hagedorn regime 
requires an entropy density higher than the 
critical Hagedorn density (of order 1 in string units).
Below this density the energy momentum tensor and hence 
the cosmology is governed by the massless 
relativistic Yang-Mills gas (the gas is
present on the brane even in the Hagedorn regime but is subdominant). 
Thus there is no problem exiting 
from the inflationary behaviour.  In fact the
main issue is how long inflation can last. We return to 
this question later when we discuss how inflation can be 
sustained.

\vspace{0.5cm}
\begin{table}[ht]
{\footnotesize 
\centerline{
\begin{tabular}{|c||c|c|c|}
\hline 
regime & 
$\rho(\beta)=E/\Vp$ & 
%$T_m^m = \eta' V_o/\Vt^2 \times ... $ & 
$- p_{bulk}$ &
$a_0(t)/a_0(0)$  \\
\hline
\hline
&&& \\
% {\bf L}[-1] & 
$d_o=0$ &
$\frac{1}{\Vt}(\beta-\beta_H)^{-2}$ & 
$\rho^{\frac{1}{2}}$ & 
$t^{\frac{4}{p}}$ \\
%{\bf L}[-$\frac{1}{2}$] & 
$d_o=1$ &
$(\beta-\beta_H)^{-\frac{3}{2}}$ & 
$\rho^{\frac{1}{3}}$ & 
$t^{\frac{6}{p}}$ \\
% {\bf L}[0] & 
$d_o=2$ &
$(\beta-\beta_H)^{-1}$ & 
$\log \rho$ & 
$\exp \left( {-C t^2 + D t }\right) $ \\
% {\bf L}[$\frac{1}{2}$] & 
$d_0=3$ &
$ (\beta-\beta_H)^{-\frac{1}{2}} $ & 
$ \rho^{-1} $ & 
$ t^{-\frac{2}{p}} $ \\
% {\bf L}[1] & 
$d_0=4$ &
$-\log (\beta-\beta_H)$ & 
$e^{- \rho}$ & 
const \\
&&& \\ \hline 
\end{tabular}
}}
\caption{Cosmological regimes for open strings in the Hagedorn phase
with $p$ large parallel dimensions, where $p=3$ for
our observable universe as a 3-brane. 
Here $d_0$ is the number of dimensions transverse to the brane
in which there are no windings; in our standard ``high-energy''
regime we take $d_0 = 0$.  Here $\rho$ is the energy density
of strings, \ie , the $T_{00}$ component of the bulk energy-momentum
tensor. In addition, $T_{55} = p_{bulk}$ is the negative bulk pressure
that drives inflation of our scale factor $a_0(t)$. Note
inflationary expansion 
for $d_o \leq 2$, i.e., at most 2 directions without windings.
The constants 
$C$ and $D$ are given for $p=3$ in eq.(\ref{expinf}).  }
%{\bf L}[$\gamma \leq 0$].}
\label{table1}
\end{table}

In all cases, we have also found
solutions to the $T_{00}$ and $T_{ii}$ equations.

\subsection{Case II:  
$\frac{\hat{\kappa}^2}
{12}\Lambda_{br}^2 - \Lambda_{bulk} =0$ and $\dot{b}\neq 0$}

We also studied the case where the extra dimension is not stabilized 
but there is still no nett cosmological constant. 
By `nett' we mean that the contribution from $a'$ and 
$n'$ in $G_5^5$ 
cancels the bulk contribution on the RHS of the $T_5^5$ equation.
The condition for this is 
\be 
\Lambda_{nett} = -\frac{\hat{\kappa}^2}{12}\Lambda_{br}^2 + \Lambda_{bulk} =0.
\ee
Using the 55 equation, we catalogued
some possible types of behavior for $a$ with different ans\"atze
for $b$.
Since $\alpha\approx 1 $, for $\gamma=-1,-\half,\half$ ($\gamma = d_o/2-1$), 
the pressure is 
\begin{equation}
\label{eq:t55}
-p_{bulk} 
%\hat{p}_\gamma 
\propto a^{{-3\gamma}\over {\gamma-1}} b^{-3\over2} \, .
\end{equation}

Taking $\Lambda_{nett}=0$, we find a family of power law 
solutions for $\gamma=-1,-\half,\half$ of the form 
\ba
\label{eq:plaw}
a_0(t) &\approx & At^q\nn\\
b_0(t) &\approx & Bt^r 
\ea 
where subscript-0 indicates values at $y=0$, 
\be
q={ {{\gamma-1}\over {2\gamma}} ({4\over 3}-r)} ,
\ee
%\ba
%q&=&{ {{\gamma-1}\over {2\gamma}} ({4\over 3}-r)},\nn\\
%A^{{3 \gamma} \over {\gamma-1}}B^{3\over 2} &=& 
%\frac{\hat{\kappa}^2}{3 q (2q-1)},
%\ea
and $r$ is arbitrary.
For our standard $\gamma = -1$,
we find superluminal solutions with $q>1$ for $r<1/3$.
Hence one can have an inflating brane with a shrinking
bulk ($r<0$) or with a growing bulk ($0<r<1/3$).
We also find a family of hyperbolic solutions with
\be
{a_o(t) \over a_o(0)} = \Biggl({{\rm sinh}2C(t+t_1) \over
{\rm sinh}2Ct_1} \Biggr)^{1/2}.
\ee
These solutions can give an
exponentially increasing scale factor on our brane.
% for
%some choices of initial value of $\dot b_o$.
The proper choice of solution in Case II depends on the
initial value of $\dot b_o$.

\subsection{Case III:
$\frac{\hat{\kappa}^2}
{12}\Lambda_{br}^2 - \Lambda_{bulk} \neq 0$ and $\dot{b}\neq 0$}

In \cite{us}
we also consider the case of additional cosmological constants
in the brane and bulk. The results can be found in our longer paper.
Not surprisingly we recover the usual cosmological constant driven
inflation when we set $\rho=0$.

\section{Sustaining inflation and solving the Horizon Problem}

In the previous sections we have found the onset of inflation
via superluminal growth of the scale factor.  We also know that
eventually, once the temperature drops out of the Hagedorn
regime, inflation ends.  This graceful exit from inflation
is an appealing feature of this scenario.  
However, we cannot calculate in between.
Once inflation begins, our calculations break down; our
interpretation of small changes in the metric corresponding
to small changes in volume no longer holds.  
At this point we can only speculate 
what happens in between.  The following are three of the possibilities:

First, previous authors have argued that strings in a de Sitter 
background are unstable to fluctuations and 
that this instability can sustain a period of de Sitter 
inflation~\cite{englert,turok,veneziano}. This phenomenon is well known 
for strings once they are in the de Sitter-like 
phase\footnote{We qualify `de Sitter' because the exponential 
solutions we have do not possess the full O(1,4) de Sitter symmetry.}. 
However the missing ingredient that the present study adds 
is an explanation for how the universe enters a de Sitter like phase 
in the first place.   Our mechanism gets the universe into
the locally de Sitter phase, whereupon the mechanism of
refs.\cite{englert,turok,veneziano} keeps it there 
{\em without} having to assume adiabaticity.

Second, our 3-brane sits in a bath of branes and bulk strings
with ongoing dynamics such as branes smashing into each other
and possibly annihilating.  The dynamics is likely to keep
the system hot.  Hence high temperatures near the Hagedorn
temperature may be sustained for a long period; inflation
continues during this period.  We note that the high
entropy of our brane $S\sim 10^{88}$ may originate from
the dynamics of this hot brane/string bath.   Eventually
the system cools, drops out of the Hagedorn regime, and inflation ends.
We are investigating this scenario in a future paper.

Third, we suggest as a possible direction for future study,
a metric in which each 4 dimensional leaf of the foliated
space is a de Sitter space.
It is well established (see for example ref.\cite{brustein} and
references therein)
that the combined (geometric plus thermal) 
system obeys a generalized second law which 
is that the total entropy,
\be 
S= S_{matter} + S_{horizon}=  S_{matter} + 
\frac{1}{4 \hat{\kappa}^2 } A_{horizon},
\ee
obeys $\delta S>0$, where $A_{horizon}$ is the area of the horizon. 
The generalized second law indicates that fluctuations 
in the geometry tend to drive the universe even further into 
the de Sitter phase provided that the Hubble constant is 
greater than the critical value.
The entropy that the horizon loses when the Hubble constant is 
increased is more than offset by the huge increase in string entropy. 
Moreover, since the specific heat of the string gas becomes larger 
when the temperature increases, inevitably
equilibrium is lost and energy flow to the strings becomes 
apparently limitless.

There is a slight difference between our study and 
the results of ref.\cite{turok} which we should comment on. 
In ref.\cite{turok}, the density never goes {\em above}
a critical density, $\rho'_c$,
and indeed asymptotes to it from below as one goes back in time. 
In our case, on the other hand, we must have $\rho > \rho_c \sim 1$ for the 
calculation to be valid (\ie to be in the Hagedorn regime).
Indeed in the present paper the energy density diverges 
as we approach the Hagedorn temperature. The difference 
is because ref.\cite{turok} was concerned with 
{\em stretched} strings and 
introduced a cut-off in the momentum integral
in order to find the fractal dimension of the string. 
This `coarse graining' put an artificial
upper limit on the amount of string that can be packed into the 
volume. In fact they found the maximum fractal dimension (\ie 2) only
at the critical density $\rho'_c$. A fractal dimension of 2 is typical of the
random walk behaviour which exists once Hagedorn behaviour sets in. 
So the coarse graining in ref.\cite{turok} 
effectively removed the Hagedorn behaviour and consequently 
ref.\cite{turok} found that
$\rho=\rho'_c$ gave $\beta=\beta_H $.
Conversely, the present paper begins in the regime $\rho > \rho_c$ 
where the calculations of ref.\cite{turok} end.

%%%%%%%%%%%%%%%%%%%%%%%%%%%%%%%%%%%%%%%%%%%%%%%%%%%%%%%%%%%%%%%%%%

\section{Conclusion and discussion}

We have studied the possible cosmological implications of
the Hagedorn regime of open strings on D-branes in the weak coupling limit. 
Our main result is that a gas of open strings can exhibit negative pressure 
leading naturally to a period of power law or even exponential inflation.
%We also find that the open string gas can dominate the cosmological 
%evolution at weak coupling even though the D-brane tension becomes
%large in this limit.
%
Hagedorn inflation also has a natural exit 
since any significant cooling can cause a change in the thermodynamics
if winding modes become  quenched or if the density drops 
below the critical density, $\rho_c\sim 1$, needed for the entropy 
of the Hagedorn phase to be dominant. 
To summarize, open strings on branes in the hot early Hagedorn phase
of the universe provide a mechanism to drive a period of
inflation, even in the absence of any potential. 
%We find this `easy-exit' feature of 
%open-string Hagedorn inflation to be of its most appealing features. 

%%%%%%%%%%%%%%%%%%%%%%%%%%%%%%%%%%%%%%%
\subsection*{Acknowledgements}

\noindent We thank Dan Chung, Cedric Deffayet, Emilian
Dudas, Keith Olive, Geraldine Servant and Carlos Savoy for discussions. 
S.A.A. and I.I.K. thank 
Jose Barb\'on and Eliezer Rabinovici for a previous collaboration
and discussions concerning this work.
S.A.A. thanks the C.E.A. Saclay for support during this work.
K.F. acknowledges support from the Department of
Energy through a grant to the University of Michigan.  K.F. thanks
CERN in Geneva, Switzerland and the Max Planck Institut fuer Physik in
Munich, Germany for hospitality during her stay. 
I.I.K. was  supported in part by PPARC rolling grant
PPA/G/O/1998/00567, the EC TMR grant FMRX-CT-96-0090 and  by the INTAS
grant RFBR - 950567. 
%%%%%%%%%%%%%%%%%%%%%%%%%%%%%%%%%%%%%%%

%%%%%%%%%%%%%%%%%%%%%%%%%%%%%%%%%%%%%%%

\end{document}